\documentclass[superscriptaddress,preprintnumbers,10pt,twocolumn,amsmath,amssymb,floats]{revtex4-1}
\usepackage{soul, ulem, todonotes}
\usepackage{amssymb,bm}
\usepackage{graphicx}
\usepackage{sidecap}
\usepackage{txfonts}
\usepackage[multi-part-units = single]{siunitx}
\usepackage{url}
\usepackage{color}
\usepackage{natbib}
\usepackage{caption}
\renewcommand{\raggedright}{\leftskip=0pt \rightskip=0pt plus 0cm}
\begin{document}
\title{Orbital-selective Dirac fermions and extremely flat bands in frustrated kagome-lattice metal CoSn}

\author{Zhonghao Liu}
\email{lzh17@mail.sim.ac.cn}
\affiliation{State Key Laboratory of Functional Materials for Informatics and Center for Excellence in Superconducting Electronics, Shanghai Institute of Microsystem and Information Technology, Chinese Academy of Sciences, Shanghai 200050, China}
\affiliation{College of Materials Science and Opto-Electronic Technology, University of Chinese Academy of Sciences, Beijing 100049, China}

\author{Man Li}
\affiliation{Department of Physics and Beijing Key Laboratory of Opto-Electronic Functional Materials$\&$Micro-Nano Devices, Renmin University of China, Beijing 100872, China}
\affiliation{Shanghai Advanced Research Institute, Chinese Academy of Sciences, Shanghai 201204, China}

\author{Qi Wang}
\affiliation{Department of Physics and Beijing Key Laboratory of Opto-Electronic Functional Materials$\&$Micro-Nano Devices, Renmin University of China, Beijing 100872, China}

\author{Guangwei Wang}
\affiliation{Department of Physics and Center for Advanced Quantum Studies, Beijing Normal University, Beijing 100875, China}

\author{Chenhaoping Wen}
\affiliation{School of Physical Science and Technology, ShanghaiTech University, Shanghai 201210, China}

\author{Kun Jiang}
\affiliation{Department of Physics, Boston College, Chestnut Hill, MA 02467, USA}

\author{Xiangle Lu}
\affiliation{State Key Laboratory of Functional Materials for Informatics and Center for Excellence in Superconducting Electronics, Shanghai Institute of Microsystem and Information Technology, Chinese Academy of Sciences, Shanghai 200050, China}
\affiliation{College of Materials Science and Opto-Electronic Technology, University of Chinese Academy of Sciences, Beijing 100049, China}

\author{Shichao Yan}
\affiliation{School of Physical Science and Technology, ShanghaiTech University, Shanghai 201210, China}

\author{Yaobo Huang}
\affiliation{Shanghai Advanced Research Institute, Chinese Academy of Sciences, Shanghai 201204, China}

\author{Dawei Shen}
\affiliation{State Key Laboratory of Functional Materials for Informatics and Center for Excellence in Superconducting Electronics, Shanghai Institute of Microsystem and Information Technology, Chinese Academy of Sciences, Shanghai 200050, China}
\affiliation{College of Materials Science and Opto-Electronic Technology, University of Chinese Academy of Sciences, Beijing 100049, China}

\author{Jiaxin Yin}
\affiliation{Laboratory for Topological Quantum Matter and Advanced Spectroscopy (B7), Department of Physics, Princeton University, Princeton, NJ 08544, USA}

\author{Ziqiang Wang}
\affiliation{Department of Physics, Boston College, Chestnut Hill, MA 02467, USA}

\author{Zhiping Yin}
\email{yinzhiping@bnu.edu.cn}
\affiliation{Department of Physics and Center for Advanced Quantum Studies, Beijing Normal University, Beijing 100875, China}

\author{Hechang Lei}
\email{hlei@ruc.edu.cn}
\affiliation{Department of Physics and Beijing Key Laboratory of Opto-Electronic Functional Materials$\&$Micro-Nano Devices, Renmin University of China, Beijing 100872, China}

\author{Shancai Wang}
\email{scw@ruc.edu.cn}
\affiliation{Department of Physics and Beijing Key Laboratory of Opto-Electronic Functional Materials$\&$Micro-Nano Devices, Renmin University of China, Beijing 100872, China}

\begin{abstract}
Layered kagome-lattice 3$d$ transition metals are emerging as an exciting platform to explore the frustrated lattice geometry and quantum topology. However, the typical kagome electronic bands, characterized by sets of the Dirac-like band capped by a phase-destructive flat band, have not been clearly observed, and their orbital physics are even less well investigated. Here, we present close-to-textbook kagome bands with orbital differentiation physics in CoSn, which can be well described by a minimal tight-binding model with single-orbital hopping in Co kagome lattice. The capping flat bands with bandwidth less than 0.2 eV run through the whole Brillouin zone, especially the bandwidth of the flat band of out-of-plane orbitals is less than 0.02 eV along $\Gamma-M$. The energy gap induced by spin-orbit interaction at the Dirac cone of out-of-plane orbitals is much smaller than that of in-plane orbitals, suggesting orbital-selective character of the Dirac fermions.
\end{abstract}
\maketitle

A typical electronic band structure of kagome lattices is characterized by a Dirac-like band capped by a flat band, which can be produced by using the tight-binding method with single-orbital nearest-neighbor hopping. Recently, the linearly dispersive energy bands and dispersionless flat bands have been partially reported in two-dimensional (2D) layered kagome-lattice 3$d$ transition metals Fe$_3$Sn$_2$  \cite{Fe3Sn2_Linda,Fe3Sn2_Lin,Fe3Sn2_Yin}, Fe$_3$GeTe$_2$ \cite{FeGeTe_Kim,FeGeTe_Zhang}, Co$_3$Sn$_2$S$_2$  \cite{CoSnS_Yin,CoSnS_Wirth,CoSnS_Wang,CoSnS_Beidenkopf,CoSnS_YLChen} and FeSn \cite{FeSn_Kang,FeSn_Lin}. Topologically protected linearly dispersive bands show distinctly different behaviors from the traditional parabolic bands. When such bands are tuned to the Fermi energy  ({$E\rm_F$}), the low-energy quasiparticle excitations would be drastically different from that of the conventional parabolic-band fermions and thus lead to novel transport properties \cite{TI_Zhang,TI_Hasan,TS_LDA}. For instance, large intrinsic anomalous Hall effects associated with Dirac/Weyl nodes near {$E\rm_F$} in antiferromagnetic (AFM) \cite{Mn3Sn_Kuroda} and ferromagnetic (FM) \cite{CoSnS_LIU,CoSnS_Wang,CoSnS_Beidenkopf,CoSnS_YLChen} kagome metals have been reported. When taking the spin-orbit coupling (SOC) into account, a small band gap can open, adding a mass term to the linearly dispersive band, and a massive Dirac fermion thus can be formed \cite{Fe3Sn2_Linda}. Contrasting with the linear band hosting massless or lightweight quasiparticles, flat band is dispersionless over a finite range of momentum, usually with super-heavy localized electrons, extremely singular density of states (DOS), and correlated states with broken symmetries.  Because of the similarity with 2D continuum Landau levels, flat bands can induce novel quantum behaviors, like Mott insulators, magnetism, fractional quantum Hall states, even at high temperatures and superconductivity  \cite{PRL_Wen,PRL_Sarma,PRL_Mudry}, as predicted in twisted bilayer graphene \cite{Gr_Cao1,Gr_Cao2}. In kagome metals, the extremely light band (Dirac fermions) and extremely heavy band (flat bands) coexist. Unconventional phenomena would happen if either of the two bands is tuned close to {$E\rm_F$}. Though the local DOS reflected by scanning tunneling spectra indicate the existence of the flat bands which have also been found in rather limited momentum space by angle-resolved photoemission spectroscopy (ARPES) in previous reported kagome materials  \cite{Fe3Sn2_Lin,CoSnS_Yin,CoSnS_Wirth,CoSnS_YLChen}, the extremely capping flat band with a much smaller bandwidth runs through the entire Brillouin zone (BZ) have not been clearly found in realistic kagome lattices.

\begin{figure*}[t!]
	\centering
	\includegraphics[width=0.8\textwidth]{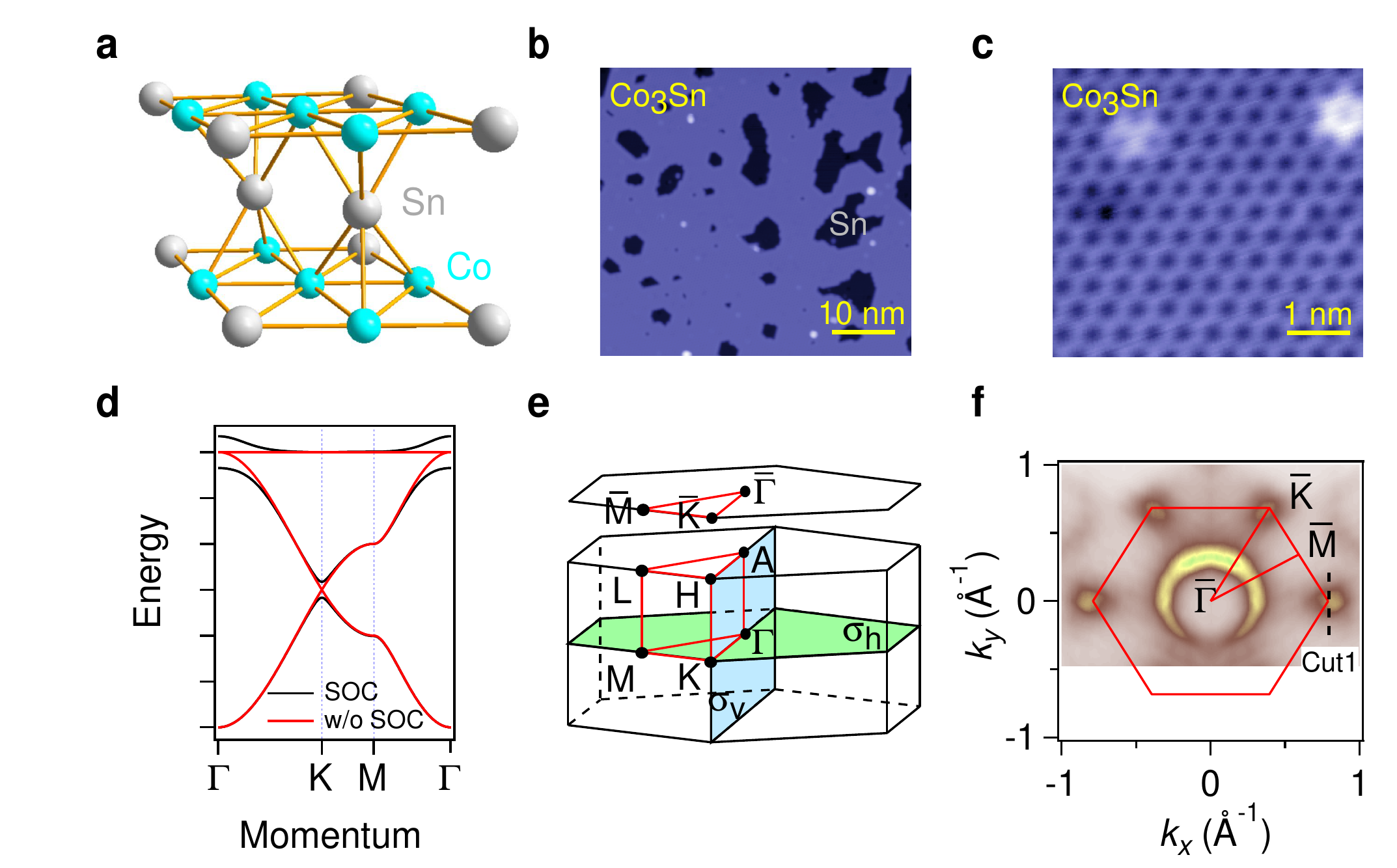}
	\caption
	{\textbf{Crystal structure, tight-binding calculations and FS.}
    \textbf{a} Crystal structure of CoSn with space group P6/mmm (No. 191).
    \textbf{b} High-resolution STM topography with $V$ = 1 V, $I$ = 20 pA. Co$_3$Sn surface (blue) and Sn surface (black) are indicated.
    \textbf{c} Atomic-resolved STM topography on Co$_3$Sn plane with $V$ = 500 mV, $I$ = 100 pA.
    \textbf{d} The typical kagome band is produced by using the tight-binding method with single-orbital nearest-neighbor hopping. The black (red) curves are indicated for with (without) the inclusion of SOC.
    \textbf{e} 3D bulk BZ with marked high-symmetry points and two mirror planes: $\sigma_h$ ($\Gamma-K-M$) and $\sigma_v$ ($\Gamma-A-H-K$). $\bar{\Gamma}-\bar{K}-\bar{M}$ plane is a projected 2D BZ.
    \textbf{f} Intensity plot at $E_F\pm$10 meV in 2D BZ. The red lines indicate high-symmetry directions and the first BZ projected on the (001) surface.
    }
	\label{1}
\end{figure*}

Moreover, five 3$d$ orbitals dominate the low-lying electronic states in these 3$d$ transition kagome materials, resulting in the presence of multiple orbital physics. Along with the increased filling of the electronic 3$d$ shell, orbital characters of the low-energy bands near {$E\rm_F$} can be tuned. Crystal field can also affect the orbital energy-levels in kagome lattice. The Dirac-like band and the flat band with certain orbital characters at {$E\rm_F$} should be associated with novel quantum phases, when the chemical potential were tuned properly. Furthermore, when SOC opens band gaps at the Dirac points, orbitals with different SOC strengths will cause various sizes of the band gaps. If the sizes of the band gaps have obvious distinctions among the orbitals, the Dirac cones with small gaps would be largely responsible for the novel electronic excitations, leading to orbital-selective Dirac fermions.  The bands with specified orbital characters have been reported in paramagnetic kagome lattice YCr$_6$Ge$_6$ ($d^4$) \cite{YCrGe_Yang}, as well as in helically coupled FM kagome layers of YMn$_6$Sn$_6$ ($d^5$) \cite{YMnSn_Li} and antiferromagnetically coupled FM kagome layers of FeSn ($d^6$) \cite{FeSn_Kang}. Mainly due to complicated magnetism, correlation effect, interlayer coupling and even the inappropriate Fermi energy, typical kagome bands with orbital differentiation physics near {$E\rm_F$} are key features that have long been theorized but still remain elusive in real kagome materials.

In this work, we clearly demonstrate the existence of sets of typical kagome bands with their orbital characters near {$E\rm_F$} in the frustrated kagome-lattice CoSn by means of ARPES, scanning tunneling microscopy/spectroscopy (STM/S), and in combination with theoretical calculations. Because of representation of the group D$\rm_{6h}$, five 3$d$ orbitals are divided into three groups: in-plane $d_{xy}$/$d_{x2-y2}$, out-of-plane $d_{xz}$/$d_{yz}$, and $d_{z2}$ orbitals, with individual sets of feature for each group. We find that the SOC induced energy gap at the Dirac cone with $d_{xy}$/$d_{x2-y2}$ orbitals characters is much larger than that with $d_{xz}$/$d_{yz}$ orbitals characters, suggesting that the latter can be predominant in the Dirac-like fermionic excitations. While, the flat bands near {$E\rm_F$} are mainly dominated by $d_{xy}$/$d_{x2-y2}$ and $d_{xz}$/$d_{yz}$ orbitals, respectively. The bandwidth of the former is less than 0.2 eV along $\Gamma-K$ and less than 0.1 eV along $\Gamma-M$. The latter, with dispersion $<$ 0.12 eV along $\Gamma-K$ and $<$ 0.02 eV which is limited by energy resolution along $\Gamma-M$, features in touch with a quadratic band at the BZ center, which is a key feature to distinguish from the flat band in heavy fermions and Mori\'e lattice. The $d_{z2}$ orbital sinks downward to higher binding energy and mainly contributes dispersions in a narrow range of momentum space around $k_z\sim\pi$ plane, indicating 2D character of the flat bands of CoSn.

\begin{figure*}[t!]
	\centering
	\includegraphics[width=0.92\textwidth]{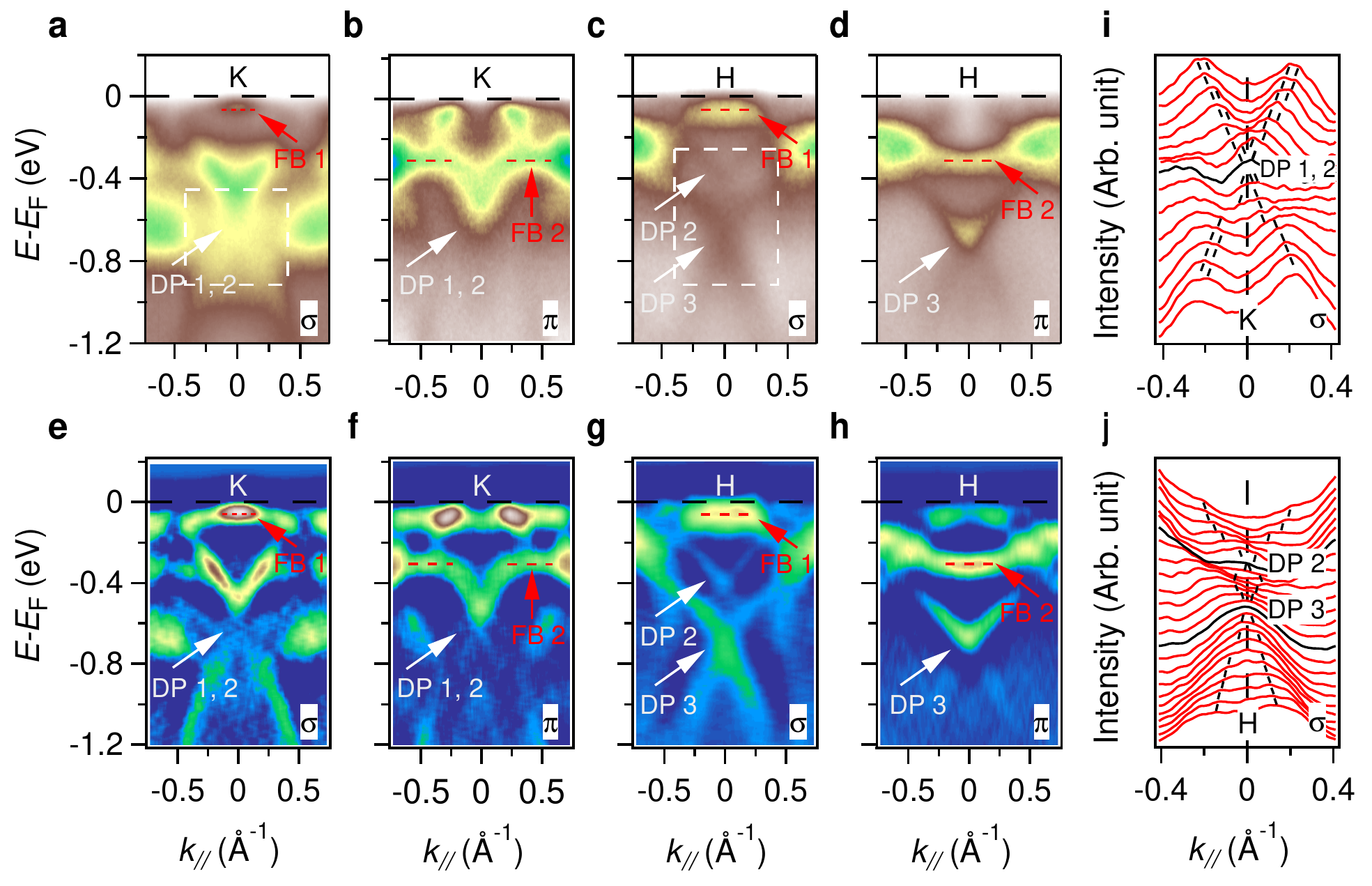}
	\caption
   {\textbf{The typical kagome bands.}
   \textbf{a}-\textbf{d} Intensity plots at the $K$ and $H$ points along cut1 as indicated in \textbf{\ref{1}f} in $\sigma$ and $\pi$ geometries, respectively. Dirac points (DP) and flat bands (FB) are indicated by the white and red arrows, respectively. The flat band top (FB 1) locates about -0.07 eV and the flat bottom (FB 2) locates about -0.3 eV below {$E\rm_F$}.
   \textbf{e}-\textbf{h} Corresponding second derivative plots of \textbf{a}-\textbf{d}.
   \textbf{i} MDCs around the Dirac cones at the $K$ point in $\sigma$ geometry show two close Dirac cones at about -0.6 eV below {$E\rm_F$}, as indicated by the dashed square in \textbf{a}. The black sticks are guides to the bands.
   \textbf{j} MDCs around the Dirac cones at the $H$ point in $\sigma$ geometry show two Dirac cones at about -0.42 eV and -0.68 eV below {$E\rm_F$}, respectively, as indicated by the dashed square in \textbf{c}.
   }
   \label{2}
\end{figure*}

\noindent\textbf{Results}

\noindent\textbf{The crystal structure and transport property of CoSn.}
CoSn is isostructural to FeSn, crystalizing in a hexagonal structure with P6/mmm (No. 191) space group, in which Co kagome lattice in single Co$_3$Sn layer is the one closest to the 2D limit, as shown in Fig. \ref{1}a.
High-quality single crystals were synthesized by the Sn flux method (see Supplementary Note 1). The samples were characterized by x-ray diffraction (XRD) (Supplementary Fig. 1) and the XRD pattern of ground crystals can be well fitted by using the reported structure of CoSn \cite{CoSnGe}. The $c$-axial electrical resistivity $\rho_{cc}$($T$) of CoSn as a function of temperature is shown in Supplementary Fig. 2, from which a clear metallic behavior is observed as the $\rho_{cc}$($T$) decreases rapidly upon cooling the sample. Further, Supplementary Fig. 3 demonstrates that CoSn is a Pauli paramagnet with very small positive magnetic moment. To directly observe the cleaved surface and kagome lattice, we performed STM experiments on the (001) surfaces of CoSn. The high-resolution STM topograph taken on the cleaved surface (Fig. \ref{1}b) shows flat Co$_3$Sn plane decorated with islands of Sn. The atomic-resolved STM topograph taken on Co$_3$Sn surface (Fig. \ref{1}c) shows perfect Co kagome lattice. With an ideal 2D kagome lattice and without complicated magnetism, the electronic bands of CoSn would be advantageous for clearly addressing the fundamental questions of kagome physics both theoretically and experimentally.

\begin{figure*}[t!]
	\centering
	\includegraphics[width=0.92\textwidth]{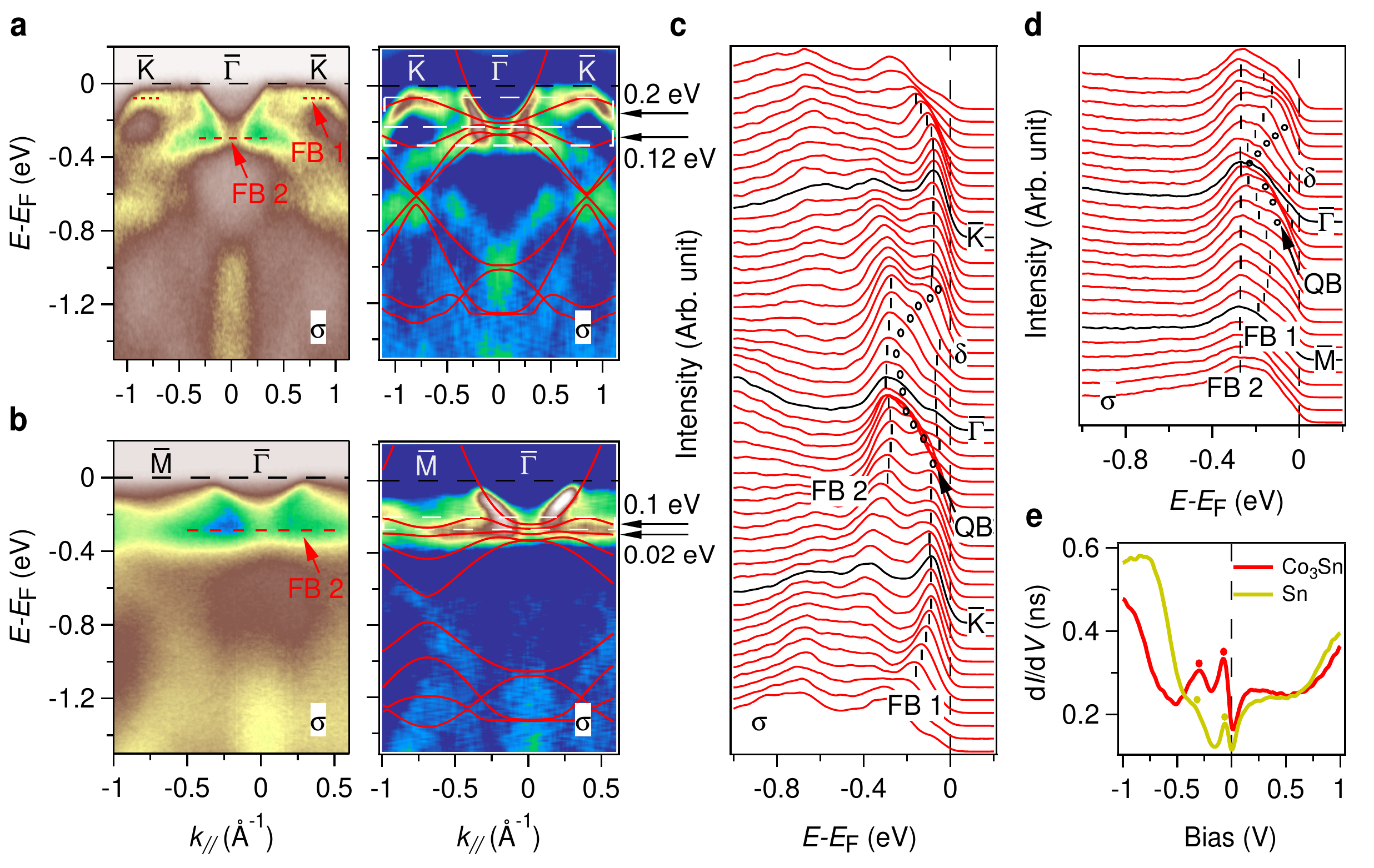}
	\caption
	{\textbf{The flat bands of CoSn.}
    \textbf{a}, \textbf{b} Intensity plots and corresponding second derivative plots along the $\bar{\Gamma}-\bar{K}$ and $\bar{\Gamma}-\bar{M}$ directions in $\sigma$  geometry, respectively. Flat bands are indicated by the red dashed lines in \textbf{a}, \textbf{b}.  The red lines on the corresponding second derivative plots are DFT calculated bands renormalized by a factor of 1.5. The bandwidths of the flat bands are marked by the white squares and indicated by the black arrows.
    \textbf{c} EDCs of \textbf{a}.
    \textbf{d} EDCs of \textbf{b}. The flat bands ( FB 1, and FB 2) and the parabolic band (QB) are indicated by the black dashed lines and open circles, respectively.
    \textbf{e} Scanning tunneling spectra (d$I$/d$V$) taken on the Co$_3$Sn surface (red) and the Sn surface (yellow). Setpoint condition: $V$ = 1 V, $I$ = 100 pA.
	}
	\label{3}
\end{figure*}

\begin{figure*}[t!]
	\centering
    \includegraphics[width=0.92\textwidth]{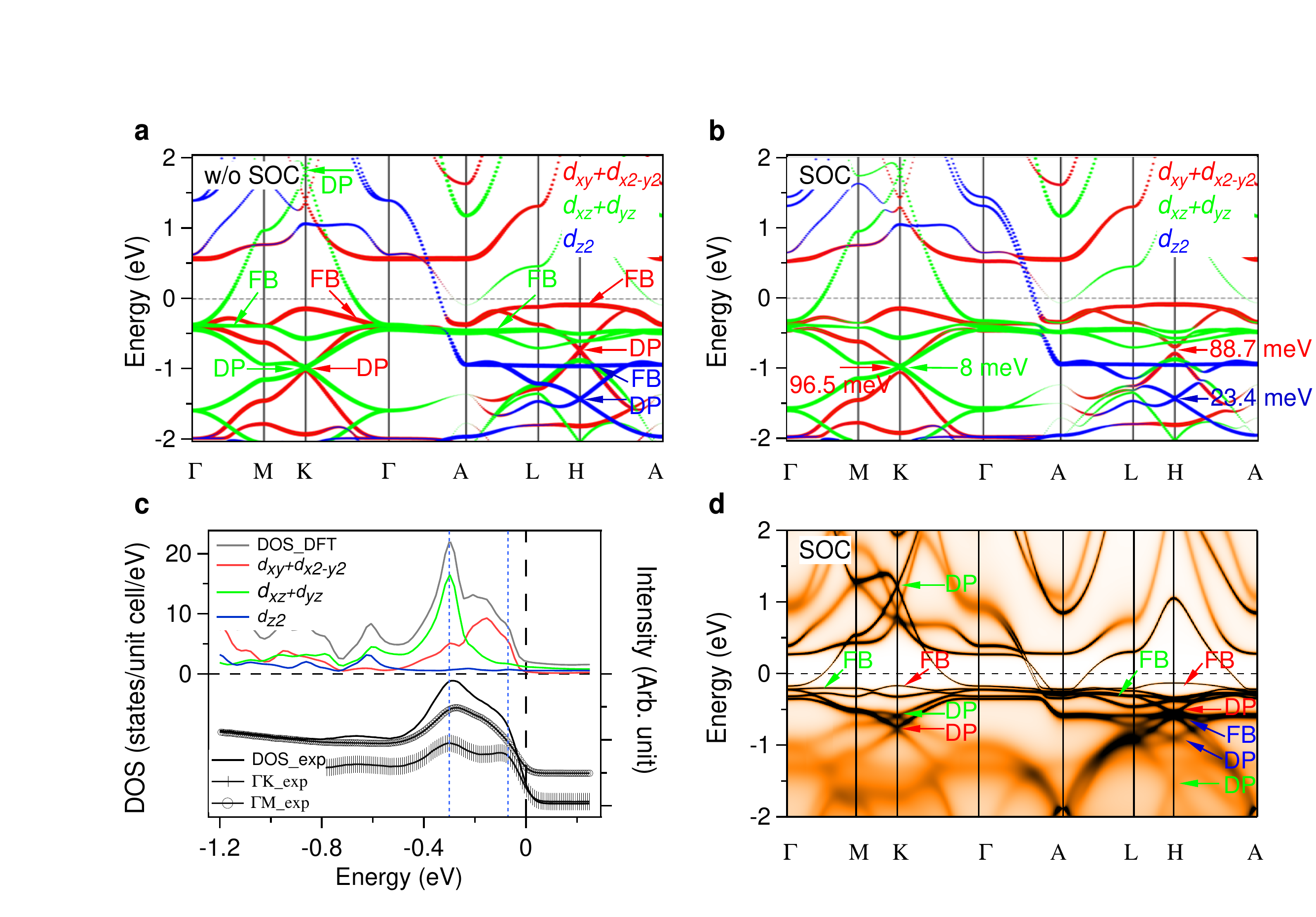}
	\caption
	{\textbf{The band calculations.}
    \textbf{a}, \textbf{b} DFT calculated bands along high-symmetry directions without and with SOC, respectively. Orbitals characters are indicated by the different colors. Flat bands (FB), Dirac points (DP), and band gaps induced by SOC at the Dirac points are indicated.
    \textbf{c} DOS from calculation and experiment. Up panel: Calculated total and orbital-resolved DOS of Co 3$d$ narrowed 1.5 times in energy range.  Down panel: Integrated EDCs in experiments.
    \textbf{d} DFT+DMFT calculation with SOC.
    }
	\label{4}
\end{figure*}

\noindent\textbf{Clear visualization of typical kagome bands of CoSn.}
The typical kagome band, characterized by a Dirac-like band capped by a flat band, is produced by using the tight-binding method with single-orbital nearest-neighbor hopping \cite{CoSnS_Yin}, as shown in Fig. \ref{1}d. When taking SOC into account, a small band gap can open at the Dirac point, and the dispersion of the flat band is modified. The bandwidth of the flat band can be further broadened by orbital and interlayer coupling in real materials. To directly detecting the features, we performed polarization- and photon-energy-dependent ARPES measurements on (001) surfaces of CoSn. The corresponding BZ with high-symmetry points and two mirror planes ($\sigma\rm_h$ and $\sigma\rm_v$) are shown in Fig. \ref{1}e. The experimental setup shows the high-symmetry directions and the normal of the sample surface defined a mirror plane (Supplementary Fig. 4). The $\pi$ ($\sigma$) geometry refers to the electric fields of the incident photons within (normal to) the mirror plane. The even (odd) orbitals with respect to the mirror plane are detected in $\pi$ ($\sigma$) geometry. With selected experimental symmetries, certain orbitals combinations, e.g. $d_{xy}$ (odd) and $d_{x2-y2}$ (even) [$d_{xz}$ (even) and $d_{yz}$ (odd)], could be enhanced or suppressed individually, thus the Dirac cones and the flat bands features could be more prominent under certain conditions.

Figs. \ref{2}a-d show ARPES intensity plots along cut1, which is perpendicular to the $\bar{\Gamma}-\bar{K}$ line, as marked on the Fermi surface (FS) plot (Fig. \ref{1}f). One can clearly see the feature of kagome bands characterized by the Dirac-like dispersion capped by the phase-destructive flat band, in spite of the intensity contrast caused by the matrix element effect under $\sigma$ and $\pi$ geometries. At $K$ point, two close Dirac cones at around -0.6 eV below {$E\rm_F$} are shown in both geometries, as shown in Figs. \ref{2}a, b, e, and f. The two close cones can be much clearly seen from the momentum distribution curves (MDCs) in Fig. \ref{2}i. The two flat bands at about -0.07 and -0.3 eV below {$E\rm_F$} are revealed in $\sigma$ (Figs. \ref{2}a, e) and $\pi$ geometries (Figs. \ref{2}b, f), respectively. The former is an extremely flat top of a hole-like band, which contributes a hotspot around the $\bar{K}$ point at {$E\rm_F$}, as shown in Fig. \ref{1}f. The latter entangles with the Dirac-like bands and looks like a break at the center of the $K$ point. At $H$ point, two Dirac cones locate at about -0.42 eV and -0.68 eV below {$E\rm_F$}, respectively, as shown in Figs. \ref{2}c, g. The two cones can also be determined from the MDCs in Fig. \ref{2}j. With the same binding energies as at the $K$ point, the two flat bands at the $H$ point are also revealed in $\sigma$ (Figs. \ref{2}c, g) and $\pi$ geometries (Figs. \ref{2}d, h), respectively, indicating that the flat bands have negligible $k_z$ dispersions. We applied the projected 2D BZ ($\bar{\Gamma}-\bar{K}-\bar{M}$) in our studies of the flat bands unless further stated, due to the weak $k_z$ dispersions and the limitation of $k_z$ resolution in ARPES experiments.

\noindent\textbf{The flat bands of CoSn.}
The band structures along the $\bar{\Gamma}-\bar{K}$ and the $\bar{\Gamma}-\bar{M}$ directions are shown in Fig. \ref{3}, and the flat bands are indicated by the red dashed lines in Figs. \ref{3}a, b. The red solid lines appended on the second derivative plots are the density functional theory (DFT) calculated bands, renormalized by a factor of $\sim$1.5 to match the main ARPES dispersive features, indicating intermediate electron correlation effects in CoSn. Along the $\bar{\Gamma}-\bar{K}$ direction, there are two phase-destructive flat bands with bandwidths of about 0.2 eV and 0.12 eV, respectively. The flat band top with the binding energy of $\sim$0.07 eV at the $\bar{K}$ point is consistent with the observation along cut1 in Fig. \ref{2}. At the $\bar{\Gamma}$ point, the electron-like bands contribute the FS around BZ center (Fig. \ref{1}f), and the nearly flat band lays at about 0.3 eV below {$E\rm_F$}. The energy distribution curves (EDCs) (Fig. \ref{3}c) clearly exhibit the flat band top (FB 1), the electron-like parabolic band (QB), and the flat band (FB 2).
As predicted by tight-binding model on kagome lattice \cite{Fe3Sn2_Yin,CoSnS_Yin}, the flat band degenerates with the quadratic band bottom at $\Gamma$ with the exclusion of SOC. The electron-like band extends to the $K$ point and forms Dirac cone above {$E\rm_F$} (see Fig. \ref{4}a). With the inclusion of SOC, the two bands further hybridize and open an energy gap of $\sim$40 meV, which can be estimated from the EDCs data. We note that the flat band in touch with the quadratic band at BZ center is a key feature to distinguish from the flat band in heavy fermions and Mori\'e lattices, but is critically missing in previous experiments. Along the $\bar{\Gamma}-\bar{M}$ direction, the two flat bands with dispersion $<$ 0.1 eV and $<$ 0.02 eV, respectively, are much narrower than that along $\bar{\Gamma}-\bar{K}$. Fig. \ref{3}d shows EDCs of Fig. \ref{3}b, clearly exhibiting the extremely flat band (FB 2) at the binding energy of $\sim$0.27 eV. As mentioned above, the clear capping flat band with such a small bandwidth through the entire BZ has not yet been reported. Complementary to ARPES measurements on the flat band of CoSn, we performed STS experiments. Fig. \ref{3}e displays the scanning tunneling spectra (d$I$/d$V$), which reflects the local DOS. Two peaks were observed around -0.07 eV and -0.3 eV on both Co$_3$Sn and Sn plane. The peak at -0.07 eV corresponds well to the flat band top at the $\bar{K}$ point measured by ARPES. The other peak reflects the flat band at about -0.3 eV below {$E\rm_F$}.

\noindent\textbf{The band calculations.}
To study the kagome bands with their orbital characters computationally, we performed DFT and dynamical mean-field theory (DMFT) calculations. Results of orbital-projected DFT calculation without and with the inclusion of SOC along high-symmetry directions are shown in Figs. \ref{4}a, b, respectively, revealing several sets of orbital-selective kagome bands. Compared with calculated and observed low-energy band structures in other 3$d$ transition AFM or FM kagome metals \cite{Mn3Sn_Kuroda,Fe3Sn2_Lin,FeGeTe_Kim,FeGeTe_Zhang,CoSnS_Wang,FeSn_Kang,FeSn_Lin}, the calculated bands of CoSn are simpler and more consistent with experimental results, mainly due to absence of complication from magnetism. Compared with calculated bands, an extra electron-like band at the $\bar{\Gamma}$ point in ARPES data (the $\delta$ band in Figs. \ref{3}c, d) should be a shadow of the band above {$E\rm_F$}, as the calculated band at the $\Gamma$ ($A$) point above {$E\rm_F$}.

On the $\Gamma-K-M$ plane, the calculated bands clear show two sets of kagome bands originating mainly from in-plane $d_{xy}$/$d_{x2-y2}$ (red) and out-of-plane $d_{xz}$/$d_{yz}$ (green) orbitals, respectively. The two cones are very close and the flat band with $d_{xz}$/$d_{yz}$ orbitals characters is more prominent along the $\Gamma-M$ lines, which are consistent well with the ARPES measurements.  Besides, the electron-like band touching with the flat bands at the $\Gamma$ point extends to the $K$ point and forms Dirac cone above {$E\rm_F$}, as discussed in Fig. \ref{3}. On the $A-H-L$ plane, the cone with $d_{xy}$/$d_{x2-y2}$ orbitals characters ascends, and the cone with $d_{xz}$/$d_{yz}$ orbitals characters descends to higher binding energy. Under the protection of time-reversal and spatial-inversion symmetry ($T$$\cdot$$P$ symmetry), the two cones at the $K$ point form two nodal lines along $k_z$ direction when the SOC is neglected (Supplementary Fig. 5). With the inclusion of SOC, the two cones open gaps with energy of $\sim$96.5 meV ($d_{xy}$/$d_{x2-y2}$) and $\sim$8 meV ($d_{xz}$/$d_{yz}$) in the DFT calculations (Fig. \ref{4}b), respectively. The strengths of SOC of in-plane orbitals are much stronger than those of out-of-plane orbitals, indicating the orbital-selective character of the SOC strength. The $d_{z2}$ (blue) orbital mainly contributes a set of kagome band at a higher binding energy around $k_z\sim\pi$ plane, showing a strong three-dimensional character.

We quantitatively compare the features between experiments and DFT calculations with SOC on the $\Gamma-K-M$ plane, details can be seen in the Supplementary Table 1. The calculated velocity ($V\rm_D$) of the Dirac cones is $\sim$1.3 times more than that in the experiment, and the relative energy shift between Dirac points and flat bands ($E_\Delta$) in calculations is $\sim$1.6 times of the experimental results. Since the flat band contributes the high DOS, the two positions of the momentum-integrated EDCs of the flat bands correspond well to DFT calculated DOS with renormalization of 1.5 in energy range, as displayed in Fig. \ref{4}c. The overall width of the peaks around -0.3 eV can be estimated as $\sim$0.2 eV. Additionally, the calculated bandwidth of in-plane orbitals is almost the same with experimental value. While, the calculated bandwidth of out-of-plane orbitals needs to be renormalized about 1.5 times along $\Gamma-K$ and 2 times along $\Gamma-M$  possibly due to the correlation effects induced by interlayer coupling. Compared to the DFT calculation, the DFT+DMFT calculated bands (Fig. \ref{4}d and Supplementary Fig. 6) are narrower by a factor of $\sim$1.5, in good agreement with experiment. The quantitative comparisons the features between DFT+DMFT results and experimental data are even much better (Supplementary Table 1), except that the bands of out-of-plane orbitals are shifted up about 0.1 eV.

\noindent\textbf{Two-orbital tight-binding model.}
Based on the above experimental and calculational findings, we propose a two-orbital tight-binding model (see Supplementary Note 4) to describe the phase-destructive flat band of $d_{xz}/d_{yz}$ orbitals. From the DFT and DMFT calculations along $k_z$ ($\Gamma-A$), we know that $d_{xz}/d_{yz}$ bands are 2D-like in the vicinity of {$E\rm_F$}. Hence, we write down an effective model inside the kagome plane. Based on D$\rm_{6h}$ point group and by including hopping integrals into second nearest neighbors \cite{Slater_Koster}, the two-orbital model with $d_{xz}/d_{yz}$ orbitals can be obtained. Supplementary Figs. 7b-d show a flat band along the $\Gamma-M-K$ line, and the degeneracy at $\Gamma$ can be lifted by introducing SOC. In contrast to the nearest-neighbor-single-band model in kagome lattice, the dispersion of the phase-destructive flat band is broadened by orbital coupling and different neighbor's hopping in the two-orbital model.

\noindent\textbf{Discussion}
We have unambiguously demonstrated the existence of sets of typical kagome bands with different orbitals characters in CoSn. The flat bands with bandwidth less than 0.2 eV run through the whole Brillouin zone, especially the bandwidth of the flat band of $d_{xz}$/$d_{yz}$ orbitals is less than 0.02 eV along $\Gamma-M$ which have yet been found in pervious reports. The flat band top at {$E\rm_F$} could induce instabilities such as superconductivity or ferromagnetism even at high temperatures, if the chemical potential were tuned properly. The band gap induced by SOC at the Dirac cone with $d_{xz}$/$d_{yz}$ orbital characters is much smaller than that with $d_{xy}$/$d_{x2-y2}$ orbital characters, suggesting orbital-selective Dirac fermions in low-energy excitations. Additionally, experimental observation of the extremely flat band over the whole BZ and in touch with the quadratic band at the BZ center is crucial in the fundamental kagome model. With an ideal 2D kagome lattice and without complication from magnetism, CoSn is a close-to-textbook example of kagome bands with interesting orbital differentiation physics, can be described by the long-sought minimal kagome tight-binding model.

\noindent\textbf{Methods}

\noindent\textbf{Sample growth and characterizations.}
High quality single crystals of CoSn were synthesized by the Sn flux method. The starting elements of Co (99.99$\%$), Sn (99.99$\%$) were put into an alumina crucible, with a molar ratio of Co : Sn = 3 : 20. The mixture was sealed in a quartz ampoule under partial argon atmosphere and heated up to 1173 K, then cooled down to 873 K with 2 K/h. The CoSn single crystals were separated from the Sn flux by using centrifuge. The samples were characterized by powder x-ray diffraction (XRD) and transport measurements, including temperature dependence of resistivity and magnetic susceptibility, and field dependence of magnetization, as shown in Supplementary Figs. 1-3.

\noindent\textbf{ARPES experiments.}
ARPES measurements were performed at Dreamline and 03U beamlines of Shanghai Synchrotron Radiation Facility (SSRF). Experimental setup is shown in Supplementary Fig. 4. Samples were cleaved $in$ $situ$, yielding a flat mirror-like (001) surface. During measurements, the temperature was kept at $T$ = 20 K. The pressure was maintained greater than $5\times10^{11}$ Torr.

\noindent\textbf{STM/S experiments.}
STM experiments were carried out with a Unisoku low-temperature STM at the base temperature of 4.3 K. CoSn single crystal samples were cleaved $in$ $situ$ under ultra-high vacuum and transferred immediately into STM head under the vacuum of 2$\times$10$^{-10}$ Torr. STS measurements were done by using standard lock-in technique with 5 mV modulation at the frequency of 914 Hz.

\noindent\textbf{Band calculations.}
DFT calculation was done in the nonmagnetic phase, whereas DFT+DMFT calculation was done in the paramagnetic phase. For DFT calculations, we used the full-potential linear augmented plane wave method implemented in Wien2K \cite{wien2k} in conjunction with Perdew-Burke-Ernzerhof generalized gradient approximation \cite{GGA} of the exchange correlation functional. DFT+DMFT was implemented on top of Wien2K and documented in Ref.~\cite{Haule_DMFT}. In the DFT+DMFT calculations, the electronic charge was computed self-consistently on DFT+DMFT density matrix. The quantum impurity problem was solved by the continuous time quantum Monte Carlo (CTQMC) method \cite{Haule_QMC,Werner}, at a temperature of 116 K and with a Hubbard $U$ = 5.0 eV and Hund's rule coupling $J$ = 0.8 eV in the paramagnetic state. The same values as we used for many iron-based compounds \cite{ZY111,ZY122,ZY11}. The experimental crystal structure (space group P6/mmm, No. 191) of CoSn with lattice constants $a$ = $b$ = 5.275 {\AA}  and $c$ = 4.263 {\AA}  was used in the calculations.

\noindent\textbf{Data availability}
The authors declare that the main data supporting the findings of this study are available within the paper and its Supplementary Information files. Extra data are available from the corresponding authors upon request.

\noindent\textbf{Acknowledgements}

This work was supported by the National Key R$\&$D Program of China (Grants No. 2016YFA0300504, 2016YFA0302300, 2018YFE0202600, 2016YFA0401000), the National Natural Science Foundation of China (Grants No. 11704394, 11774423, 11774421, 11822412, 11674030, 11574394, U1832102), the Fundamental Research Funds for the Central Universities, and the Research Funds of Renmin University of China (RUC) (15XNLQ07, 18XNLG14, 19XNLG17). C.W. and S.Y. acknowledge the financial support from Science and Technology Commission of Shanghai Municipality (STCSM) (Grant No. 18QA1403100). K.J. and Z.W. are supported by the U.S. Department of Energy, Basic Energy Sciences Grant No. DE- FG02-99ER45747. The ARPES experiments were performed on the Dreamline beamline of SSRF and supported by the Ministry of Science and Technology of China (2016YFA0401002) and the CAS Pioneer Hundred Talents Program. Part of this research used Beamline 03U of the SSRF, which is supported by ME2 project (11227902) from NSFC. The calculations used high performance computing clusters of Beijing Normal University in Zhuhai and the National Supercomputer Center in Guangzhou.

\noindent\textbf{Author contributions}

Z.L., Z.Y., H.L., and S.W. provided strategy and advice for the research. Z.L., M.L., X.L., Y.H., D.S., and S.W. performed ARPES measurements; C.W., and S.Y. performed STM measurements; G.W., and Z.Y. carried out DFT and DFT+DMFT calculations; K.J., J.Y., and Z.W. proposed theoretical model; Q.W., and H.L. synthesized the single crystals. All authors contributed to writing the manuscript.

\noindent\textbf{Competing interests}

The authors declare no competing interests.

\normalem
\bibliography{Biblio}
\end{document}